\documentclass[journal=arxiv,manuscript=article]{achemso}

\usepackage[version=3]{mhchem} 
\usepackage{graphicx}

\author{Raphael M. Tromer}
\author{Levi C. Felix}
\affiliation[State University of Campinas]
{Applied Physics Department, State University of Campinas, Campinas, SP, 13083-970, Brazil}
\alsoaffiliation[State University of Campinas]
{Center for Computational Engineering and Sciences, State University of Campinas, Campinas, SP, 13083-970, Brazil}
\author{Cristiano F. Woellner}
\affiliation[Federal University of Parana]
{Physics Department, Federal University of Parana, UFPR, Curitiba, PR, 81531-980, Brazil}
\author{Douglas S. Galvao}
\email{galvao@ifi.unicamp.br}
\affiliation[State University of Campinas]
{Applied Physics Department, State University of Campinas, Campinas, SP, 13083-970, Brazil}

\title{On the Structural Stability and Optical Properties of Germanium-based Schwarzites: A Density Functional Theory Investigation}


\begin{document}

\begin{tocentry}

Some journals require a graphical entry for the Table of Contents.
This should be laid out ``print ready'' so that the sizing of the
text is correct.

Inside the \texttt{tocentry} environment, the font used is Helvetica
8\,pt, as required by \emph{Journal of the American Chemical
Society}.

The surrounding frame is 9\,cm by 3.5\,cm, which is the maximum
permitted for  \emph{Journal of the American Chemical Society}
graphical table of content entries. The box will not resize if the
content is too big: instead it will overflow the edge of the box.

This box and the associated title will always be printed on a
separate page at the end of the document.

\end{tocentry}

\begin{abstract}
  Since graphene was synthesized the interest for building new 2D and 3D structures based on the carbon allotropes has been growing every day. One of these 3D structures is know as carbon schwarzites. Schwarzites consist of carbon nanostructures possessing the shape of Triply-Periodic Minimal Surfaces (TPMS), which is characterized by a negative Gaussian curvature introduced by the presence of carbon rings with more than six atoms. Some examples of schwarzite families include: primitive (P), gyroid (G) and diamond (D). Previous studies considering different element species of schwarzites have investigated the mechanical, electrical and thermal properties. In this work, we investigated the stability of germanium (Ge) schwarzites using density functional theory with GGA exchange-correlation functional. We chose one structure of each family (P8bal), (G688) and (D688). It was observed that regions usually flat in carbon schwarzites acquires buckled configurations as previously observed on silicene and germanene monolayers. The investigated structures presented a semiconducting bandgap ranging from $0.13$ to $0.27$ eV. We also performed calculations of optical properties within the linear regime, where it was shown that Ge schwarzites structures absorb light from infrared to ultra-violet frequencies. Therefore, our results open new perspectives of materials that can be used in optelectronics devices application.
\end{abstract}

\section{Introduction}

Proposed in 1991 by Mackay and Terrones\cite{mackay_1991}, schwarzites are crystalline structures where triply periodic minimal surfaces (TPMS) are decorated with carbon atoms along their surface. Although schwarzites have not been synthesized yet, they were found to be energetically stable \cite{terrones_1992,okeeffe_1992,terrones_1997,vanderbilt_1992}, thus, motivating the recent investigations on possible synthetic routes, such as zeolite templating\cite{nishihara_2009,nishihara_2018,braun_2018} and multifold C-C coupling reactions\cite{farrell_2019}.

Theoretical investigations have shown that schwarzites possess unique mechanical properties, such as the ability to be compressed to very high strain values without breaking\cite{sajadi_2018,miller_2016,felix_2019,jung_2018}, high energy-absorption performance\cite{felix_2019,pedrielli_2017}, near-zero Poisson's ratio for most structures \cite{felix_2019,miller_2016} and an auxetic behavior (negative Poisson's ration) only in a range of deformation for some structures\cite{felix_2019}. Also, many potential applications have been proposed, such as catalysis, molecular sieving\cite{lu_1996,kyotani_2000}, gas storage\cite{damascenoborges_2018,Collins2019}, alkali ion batteries\cite{park_2010}, as anode for lithium-ion batteries\cite{odkhuu_2014} and energy-absorbing materials\cite{barborini_2002,benedek_2003,wu_2015,townsend_1992,pedrielli_2017}. Different schwarzite structures possess distinct electronic properties, where the bandgap values vary from metallic to semiconductor\cite{valencia_2003,phillips_1992,gaito_1998,huang_1993,tagami_2014}. Interestingly, first-principles calculations\cite{weng_2015} predict that some structures can even exhibit Dirac-like points similar to graphene. Schwarzites have also interesting magnetic properties, as some structures were predicted to present a net magnetic moment in their electronic ground state\cite{park_2003}. It has also been predicted that the presence of negative Gaussian curvatures can induce suppression in the lattice thermal conductivity\cite{zhang_2017,pereira_2013}, which can be further tuned by the introduction of guest atoms in their pores making them good candidates for thermoelectric applications\cite{zhang_2018}.

Despite being originally proposed as carbon-based materials, in principle schwarzites can also be made of other elements and in fact, this has been already investigated. Boron nitride-based (BN) schwarzites were found to be wide bandgap semiconductors or insulators depending on the structure topology \cite{gao_2017}. Also, hydrides and oxides of boron, carbon, nitrogen, aluminum,\cite{laviolette_2000} and silicon\cite{zhang_2010} have been investigated with density functional theory. Silicon (Si) and germanium (Ge) have a stronger sp$^3$ character than carbon and significant structural differences are expected to occur in many analogs of carbon nanostructures such as closed cage structures\cite{rthlisberger_1994}, nanotubes\cite{zhang_2002,durgun_2005,bai_2004,guo_2012,fagan_2000,singh_2004}, and even two-dimensional honeycomb graphene-like sheets\cite{takeda_1994,cahangirov_2009,topsakal_2010,zhao_2012,botari_2014}, the so-called silicene and germanene, which have already been experimentally realized\cite{ni_2012,ohare_2012,vogt_2012,dvila_2014}. More recently, multilayer silicene has been synthesized and shown to be promising for electronic applications \cite{vogt_2014} as well as stable under ambient conditions\cite{depadova_2014}. These important and significant differences among carbon and silicon and/or germanium structures appear mainly in the form of structural buckling due to the pseudo-Jahn–Teller effect (PJTE)\cite{bersuker_2013,hobey_1965}. Structural buckling of both silicene and germanene structures have been already demonstrated\cite{cahangirov_2009}. Due to these differences, unique silicon and germanium structures can exist with no corresponding carbon counterpart\cite{perim_2014}. Based on that it would be interesting to investigate the structural and electronic properties of germanium-based schwarzites. Schwarzite families contain many structures, in the present work we select one representative structure of the primitive (P8bal), gyroid (G688), and diamond (D688) families (see Figure \ref{fig:optimized}).

\section{Materials and Methods}

In order to investigate the structural stability of germanium schwarzites, 
we used first-principles calculations based on density functional theory (DFT),
as implemented in the SIESTA code \cite{Soler_2002}. The exchange-correlation term is given in the generalized gradient approximation (GGA-PBE) \cite{Perdew_1996}. The wave functions are described as a linear combination of atomic
orbitals with a sum of z-basis set with polarized function (DZP) \cite{Soler_2002}. The norm-conserving Troullier-Martins pseudopotential (with a Bylander factorized form) is used to represent the  interaction between valence electrons and atomic ions \cite{Troullier_1993,Kleinman_1982}.

The reciprocal space is sampled on a Monkhorst-Pack scheme with a  $2\times 2 \times2$ k-point mesh \cite{Monkhorst_1976}. The calculations were performed with a mesh cut-off of $200$ Ry considering that the Brillouin zone is sampled by k-points along with a simple cubic cell. 

We considered that each self-consistent calculation cycle is performed until the convergence is achieved when the maximum difference between elements of the density matrix is smaller than $10^{-4}$ eV. In the structural optimization procedure, we let the atoms and lattice parameters vary until the convergence is achieved when the force on each atom was less than $0.01$ eV\AA.

Then we computed the formation energy for each schwarzite structure
composed of germanium atoms and for comparison purposes, we also considered
germanene and germanium at fcc unit-cell (diamond structure).

In order to investigate if occur appreciable structural changes at room temperature, we performed ab initio
molecular dynamics (AIMD) simulations using the siesta code at $T=300$ K. We used a
time step of $0.1$ fs in an NVT ensemble for a total simulation time of $2$ ps.
For control the temperature in the AIMD simulations we employed the
Nosé-Hoover thermostat.

From the optimized structures, we performed optical calculations using the
same parameters earlier discussed (used in siesta code). The optical properties
were obtained assuming that an external electric field of magnitude $1.0$ V/\AA~
is polarized as a medium of the three spatial directions.

Trough the complex dielectric function $\epsilon =\epsilon_1+i\epsilon_2$,
were determined all relevant optical properties, where $\epsilon_1$ and $\epsilon_2$ are real
and imaginary, respectively.

From Kramers-Kronig transformation, is possible to derive $\epsilon_1$ as

\begin{equation}
\epsilon_1(\omega)=1+\frac{1}{\pi}P\displaystyle\int_{0}^{\infty}d\omega'\frac{\omega'\epsilon_2(\omega')}{\omega'^2-\omega^2},
\end{equation}
where $\omega$ is the frequency of photon.

From Fermi's golden rule, its possible to show that the imaginary part, $\epsilon_2$, is given by:

\begin{equation}
\epsilon_2(\omega)=\frac{4\pi^2}{\Omega\omega^2}\displaystyle\sum_{i\in \mathrm{VB},j\in \mathrm{CB}}\displaystyle\sum_{k}W_k|\rho_{ij}|^2\delta	(\epsilon_{kj}-\epsilon_{ki}-\omega),
\end{equation}

where VB and CB denotes the valence and conduction band, respectively, $\Omega$ is the unit
cell volume and $\rho_{ij}$ is the dipole transition matrix element.

Once the part real and imaginary of dielectric function were determined, the absorption
coefficient $\alpha$, reflectivity $R$ and refractive index $\eta$, are determined directly by expressions:

\begin{equation}
\alpha (\omega )=\sqrt{2}\omega\bigg[(\epsilon_1^2(\omega)+\epsilon_2^2(\omega))^{1/2}-\epsilon_1(\omega)\bigg ]^{1/2},
\end{equation}
\begin{equation}
R(\omega)=\bigg [\frac{(\epsilon_1(\omega)+i\epsilon_2(\omega))^{1/2}-1}{(\epsilon_1(\omega)+i\epsilon_2(\omega))^{1/2}+1}\bigg ]^2 ,
\end{equation}
\begin{equation}
\eta(\omega)= \frac{1}{\sqrt{2}} \bigg [(\epsilon_1^2(\omega)+\epsilon_2^2(\omega))^{1/2}+\epsilon_1(\omega)\bigg ]^{2}.
\end{equation}

\section{Results and discussion}
\subsection{Structural stability at $T=0$ K}
\begin{figure}
\centering
\includegraphics[width=0.42\linewidth]{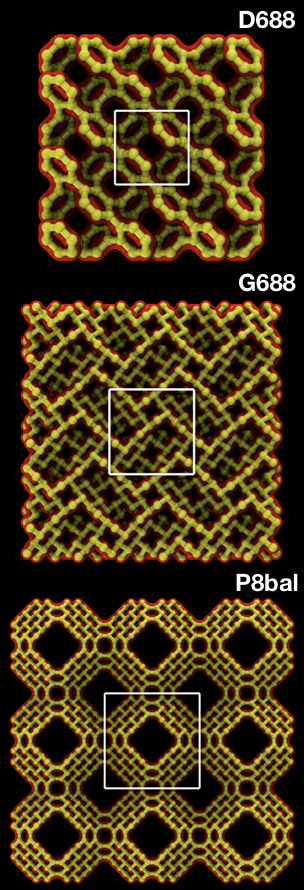}
\caption{Optimized structures of germanium-based schwarzites with square unit cell highlight replicated $2\times 2\times 2$: a) D688, b) G688 and c) P8bal. In the inset of figure \ref{fig:optimized}-c) we can see the observed pronounced buckling.}
\label{fig:optimized}
\end{figure}

In Figure \ref{fig:optimized} we present the optimized germanium-based schwarzite structures. All structures are stable at $T=0$ K. D688, G688 and P8bal contain 24, 96 and 192 atoms in their unit cell, respectively. The obtained structural shapes are very similar to the ones of carbon-based schwarzites, with the exception of the buckling (see inset of figure \ref{fig:optimized}-c)), which becomes more pronounced as the structures contain larger flat segments and is a consequence of the pseudo-Jahn-Teller effect \cite{bersuker_2013,hobey_1965}. Another difference is the typical bond length values, which are $\sim$ $2.4$ \AA~ for germanium in comparison to $\sim$ $1.5$ \AA~ for carbon \cite{Feng_2020}.   

We also calculated the formation energy of the schwarzites and
compared them to the corresponding values of germanene (already synthesized) and germanium (diamond structure) calculated using the same computational approach.
The formation energy values were obtained using the following expression:
\begin{equation}
E_f=\frac{E_{\rm structure}-N_{\rm Ge}E_{\rm Ge}}{N_{\rm Ge}},
\label{eq:formation}
\end{equation}
where $E_{\rm structure}$ is the total energy of the structure, $E_{\rm Ge}$ is the total energy of one isolated germanium atom and $N_{\rm Ge}$ is the number of
germanium atoms present in each structure. We calculated the total energy of a germanium atom considering it placed within a large simulation box.

The results are presented in Table \ref{tab:formation}.
We can see that the energies for schwarzites are not very high in comparison to other structures occurring in nature or/and that were previously synthesized. This is indicative of their synthesis feasibility.
\begin{table}[]
    \centering
    \begin{tabular}{c|c}
    \hline
    \hline
    Structure     & $E_{\rm f}$ (eV/atom)  \\
    \hline
    D688     & -4.25\\
    \hline
    G688  &  -4.39 \\
    \hline
    P8bal & -4.30 \\
    \hline
    Germanium  (fcc) & -4.63 \\
    \hline
    Germanene & -4.40 \\
    \hline
    \hline
    \end{tabular}
    \caption{Formation energies of germanium-based schwarzites, germanene, and germanium (fcc).}
    \label{tab:formation}
\end{table}

In order to verify whether significant structural changes occur at room temperature, we also performed AIMD with a NVT ensemble during $2$~ps at $T=300$~K.

\begin{figure}
\begin{center}
\includegraphics[width=0.8\linewidth]{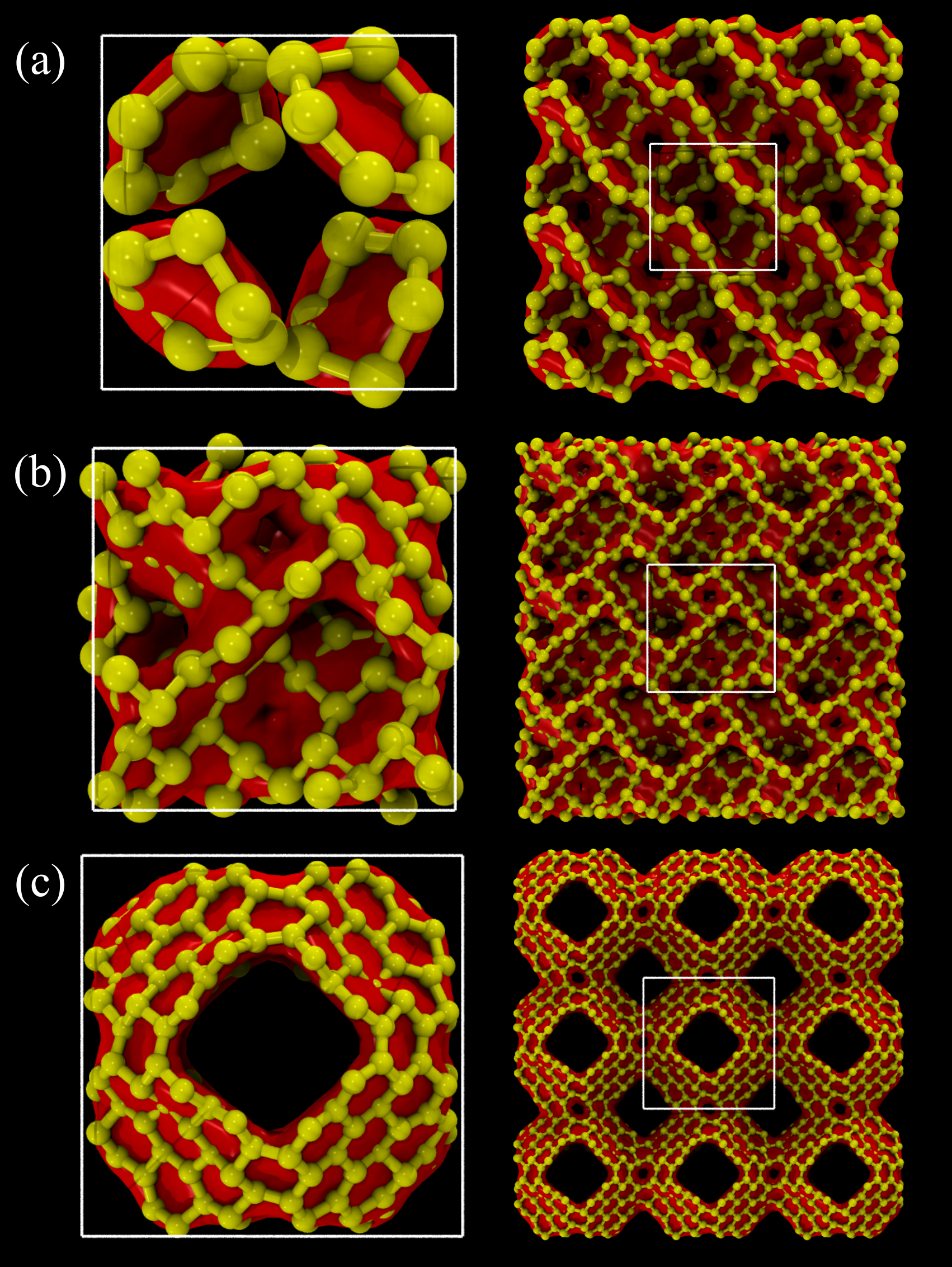}
\caption{Snapshots of the obtained equilibrated structures after $2$~ps AIMD simulations. (a) D688; (b) G688; and (c) P8bal. Left and right correspond to the used unit cells and their replicated ($3\times 3\times 3$) supercells, respectively.}
\label{fig:md}
\end{center}
\end{figure}

In Figure \ref{fig:md} we present snapshots of the obtained equilibrated structures after $2$~ps AIMD simulations. When the thermostat is set on, the shape of structures remained similar to those obtained at $T=0$~K and showed in the figure \ref{fig:optimized}. There are only small structural differences between the two cases, which can be attributed to thermal fluctuations. These results indicate that the structures are stable at room temperature and do not exhibit significant thermal induced structural changes.

\subsection{Electronic Analysis}

\begin{figure}
\begin{center}
\includegraphics[width=0.6\linewidth]{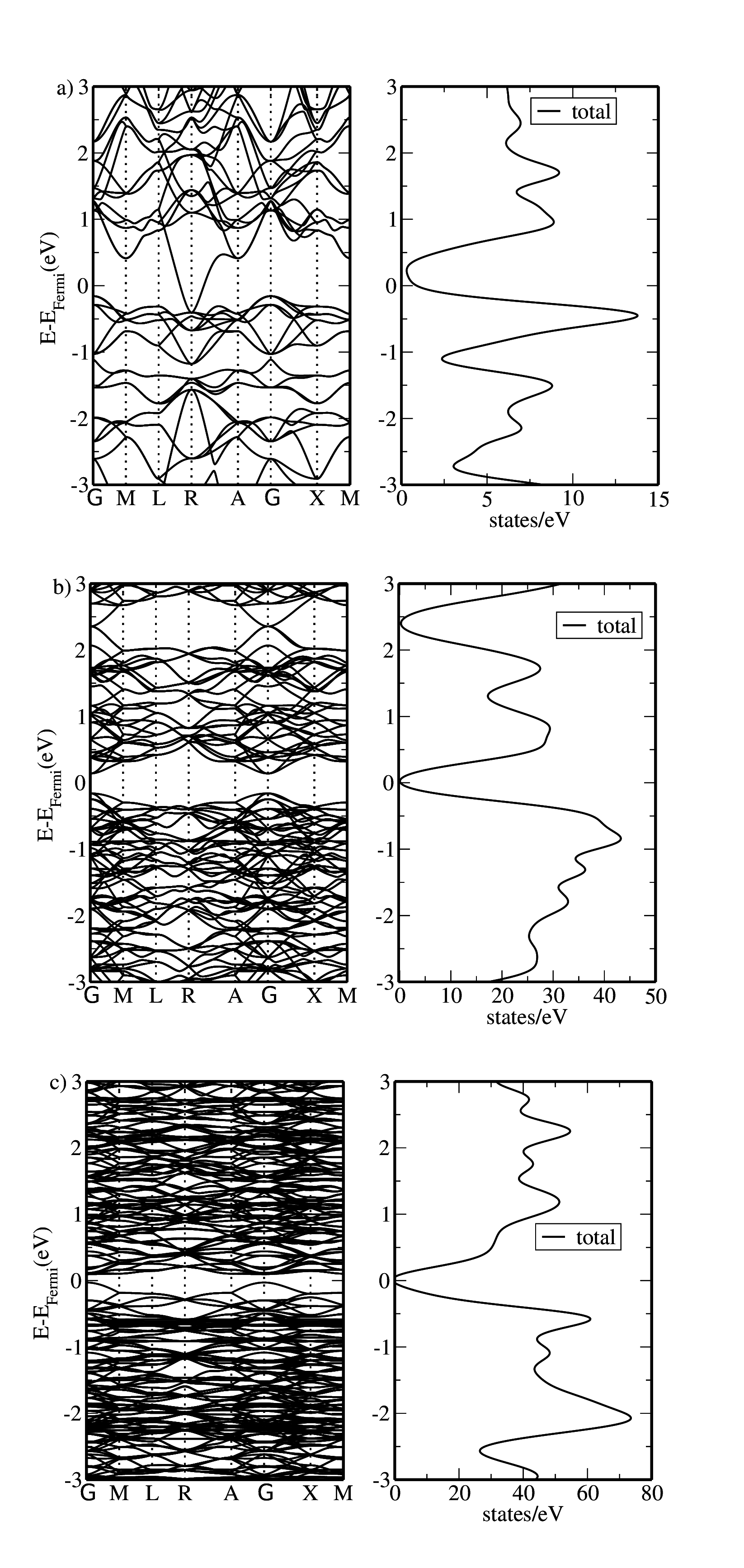}
\caption{Calculated electronic band structures and their corresponding projected total density of states for the schwarzites shown in \ref{fig:optimized}. The Fermi level value is set to zero.}
\label{fig:bandas}
\end{center}
\end{figure}

In Figure \ref{fig:bandas} we present the calculated electronic band structures and their corresponding projected total density of states (PDOS) for the schwarzites shown in \ref{fig:optimized}. The displayed band structures used the same special k-points from reference \cite{Feng_2020}.

From Figure \ref{fig:bandas} we can see that only the D688 presents a metallic behavior, the valence/conduction bands touch at R symmetry point. The structures G688 and P8bal are semiconductors with a small direct gap at $\Gamma$ point with values of $0.27$ and $0.13$ eV, respectively.

It is interesting to notice that although carbon and germanium are in the same IV group and the schwarzite structures are quite similar (with the exception of a more pronounced buckling in the case of germanium), their electronic structures are quite different, especially with relation to the bandgap values.  Carbon-based D66, G688 and P8bal schwarzites present bandgap values of $2.9$, $1.5$, $1.4$~eV, respectively \cite{valencia_2003}, i. e., much larger values than the corresponding germanium structures. 

\begin{figure}
\begin{center}
\includegraphics[width=1.0\linewidth]{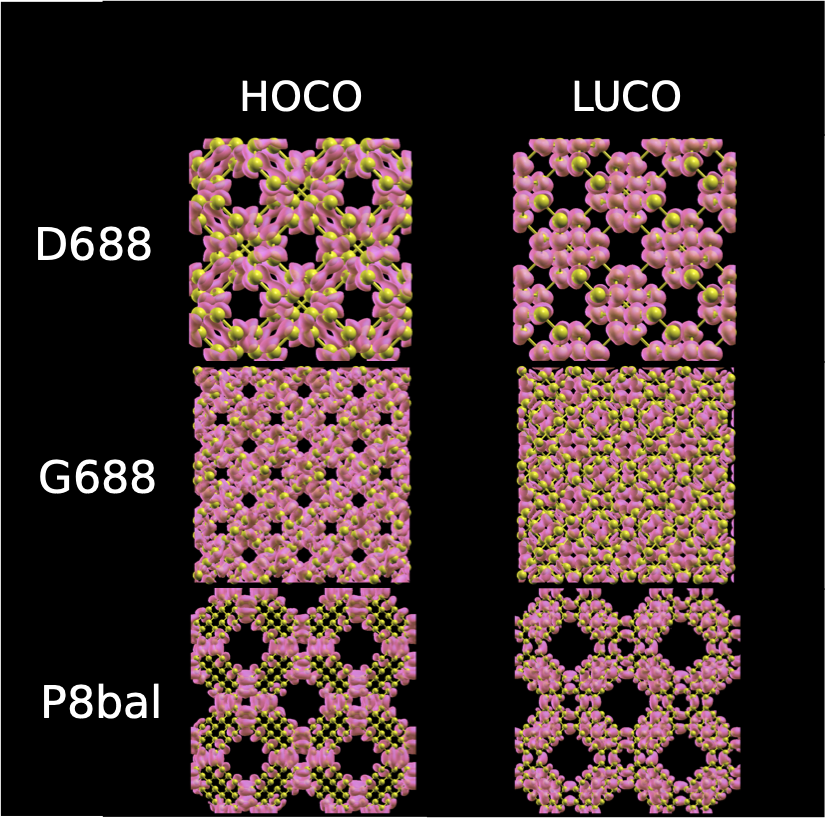}
\caption{HOCO (left) and LUCO (right) of the schwarzites shown in Fig. \ref{fig:optimized}.}
\label{fig:orbitals}
\end{center}
\end{figure}

In Figure \ref{fig:orbitals} we show the highest occupied crystal orbital (HOCO) and lowest-unoccupied crystal orbital (LUCO) of the schwarzites shown in \ref{fig:optimized}. We can see from this Fig. that the HOCO and LUCO are well delocalized for all structures.

\subsection{Optical Analysis}

\begin{figure}
\begin{center}
\includegraphics[width=1.0\linewidth]{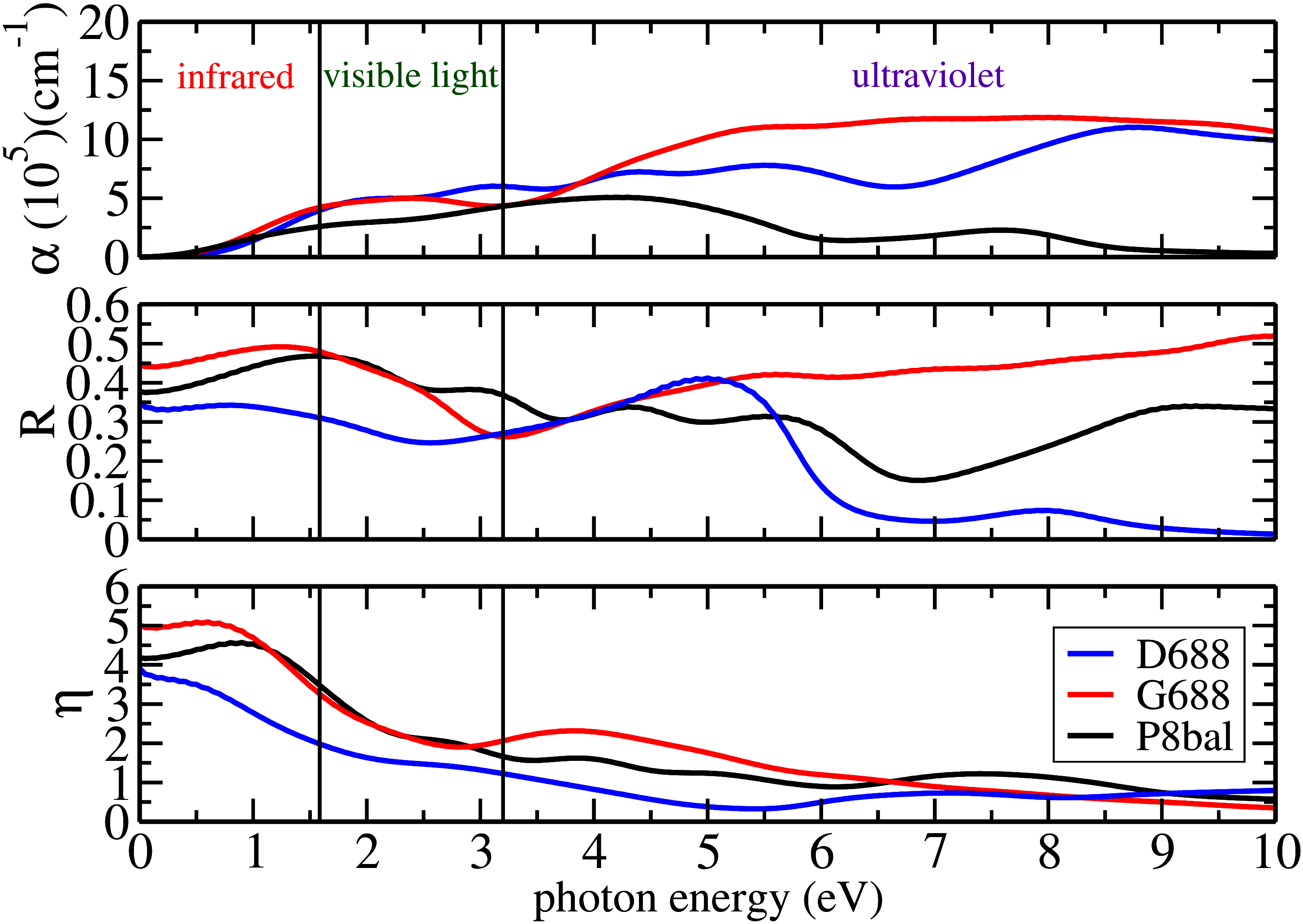}
\caption{From top to bottom: (a) absorption coefficient; (b) reflectivity; and (c) refractivity index as a funtion photon energy for schwarzites shown in Fig. \ref{fig:optimized}.}
\label{fig:optical}
\end{center}
\end{figure}

In Fig. \ref{fig:optical} we present the optical coefficients results as a function of photon energy value. The light is polarized as the average of $x$, $y$ and $z$ directions. In the Fig. \ref{fig:optical}-a), we present the absorption coefficients results. The absorption starts in the infrared region near zero for all cases. This is expected because D688 is metallic and G688 and P8bal are semiconductors with small direct bandgap values ($0.27$ and $0.13$ eV, respectively at $\Gamma$ point). These bandgaps are associated with the first optical transition between HOCO to LUCO. The analyses of the PDOS (\ref{fig:bandas}) show that these transitions involve bonding ($p_{x,y,z},\pi$) orbitals to anti-bonding ($p_{x,y,z},\pi^*$) ones. There is no preferential polarization axis because of the spatial orbital symmetries.

We noticed that all germanium-based schwarzite structures have a small  optical activity in infrared region. The same behavior was observed for germanene \cite{JOHN_2017}, germanium (diamond) \cite{Eunice_2010} and germanium-doped graphene structures \cite{ger_2017}. The structures have almost the same behavior in the infrared regions and become more differentiated for the visible and ultraviolet regions.

In Fig. \ref{fig:optical}-b and c we present the results for reflectivity and refractivity, respectively. All structures exhibit good reflectivity in the infrared, visible and ultraviolet (with the exception of D688). The corresponding refractivity continuously decreases from infrared to ultraviolet regions. Therefore the major part of the incident light will be absorbed and a small quantity will be reflected. This is indicative that the structures could be good candidates for solar cell applications, ultraviolet blocks and/or other optoelectronic devices.

\subsection{Mechanical Properties}

\begin{figure}
\begin{center}
\includegraphics[width=0.60\linewidth]{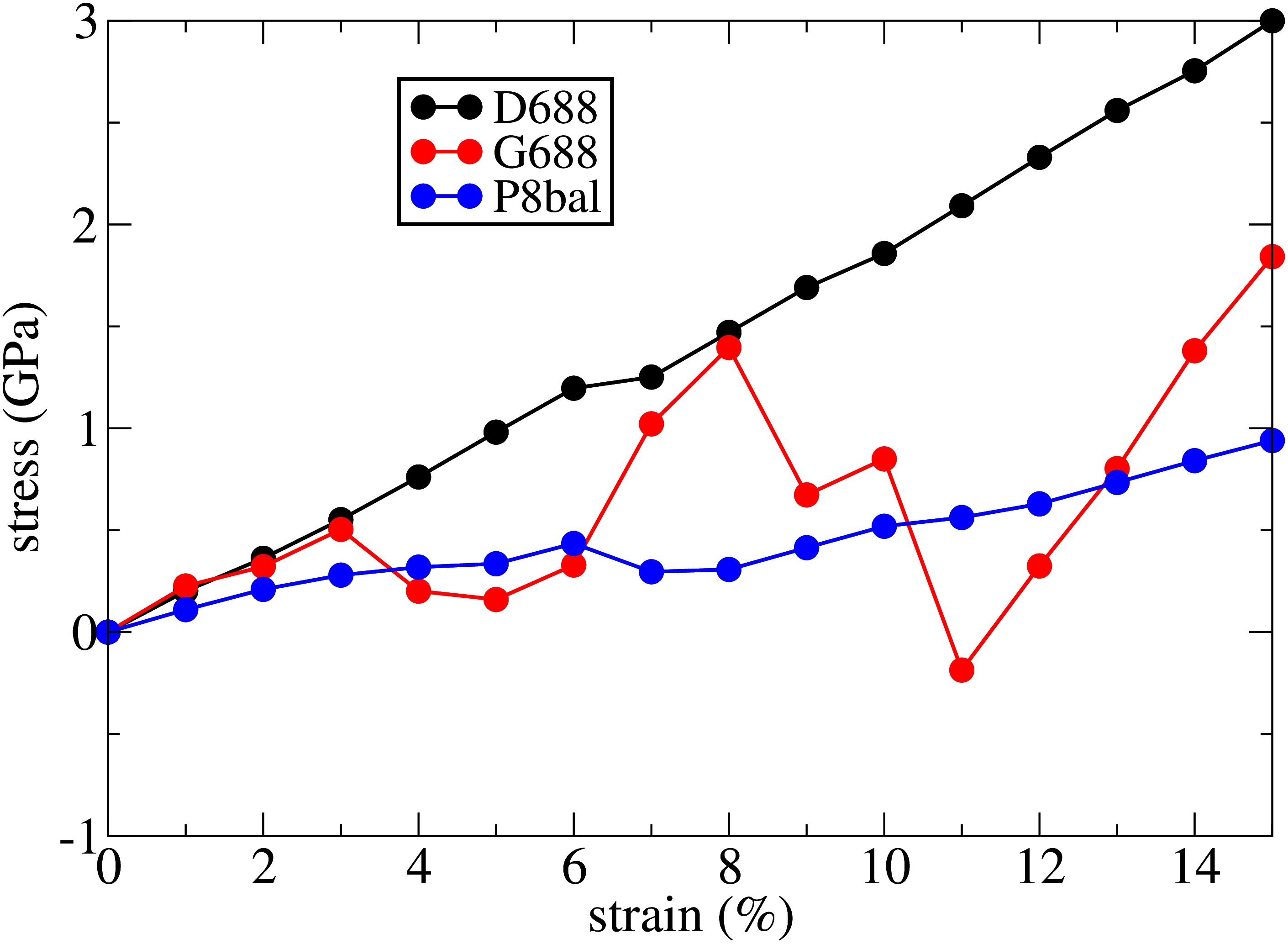}\\
\includegraphics[width=0.58\linewidth]{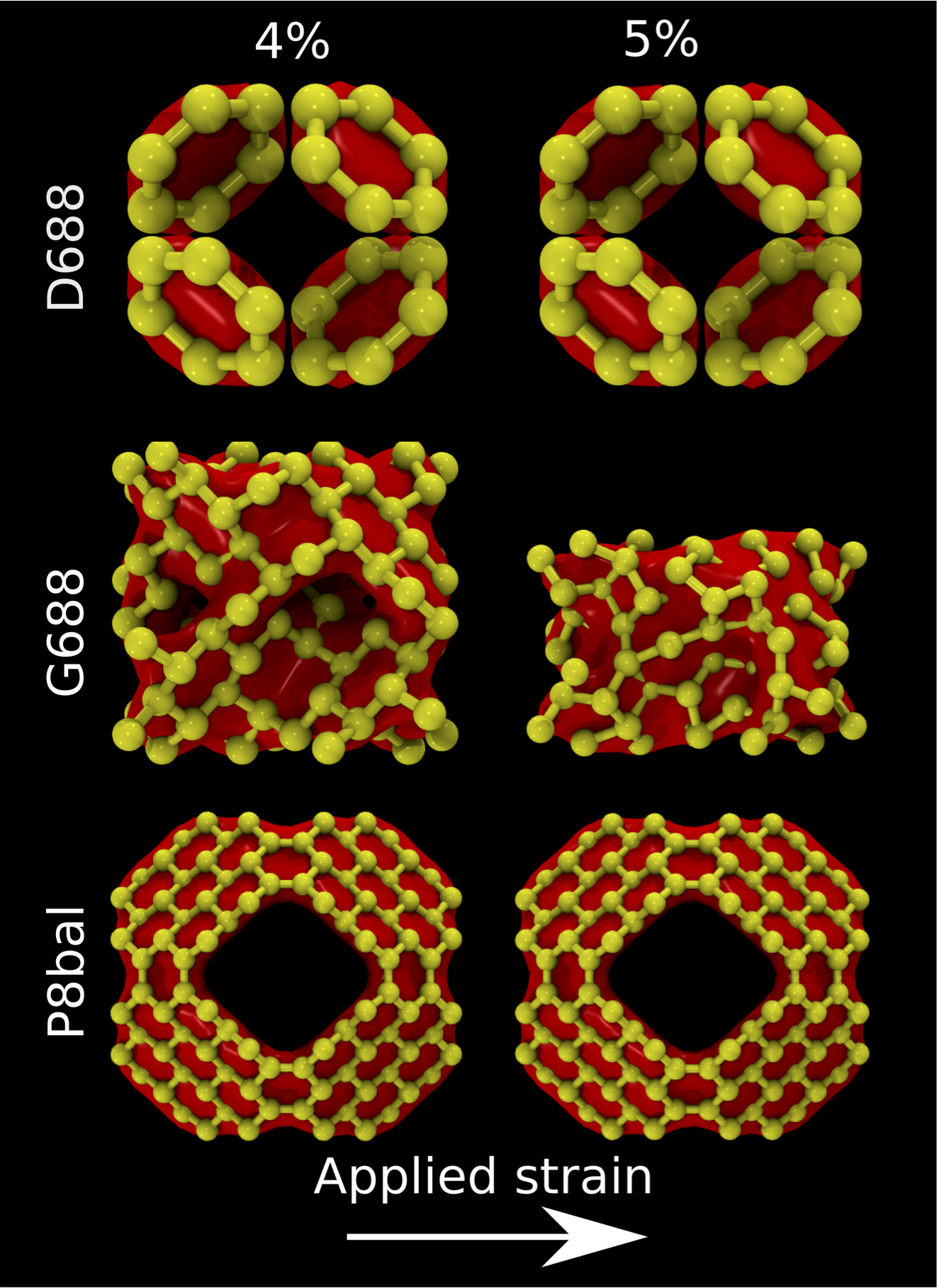}
\caption{Top: stress x strain curves for the structures shown in Fig. \ref{fig:optimized}; Bottom: representative snapshots of the deformed structures at 4 and 5\%, respectively.}
\label{fig:stress}
\end{center}
\end{figure}

In Fig. \ref{fig:stress}-top) we present the strain-stresss curves for the structures shown in Fig. \ref{fig:optimized} considering an applied uniaxial tensile deformation along the x-crystal axis. The Young's modulus values can be estimated from the linear region of the curves. The obtained values are $20$, $22$ and $11$~GPa for D688, G688, and P8bal, respectively. These values are typically one order of magnitude lower than the corresponding carbon-based structures ( $139$, $156$, and $67$ \cite{miller_2016}), but preserved the ordering values. 

The structures D688 and P8bal exhibit similar trends in the stress-strain curves, while G688 a quite different one. This can be explained by the different deformation mechanisms. In the bottom part of the Fig. \ref{fig:stress} we present representative snapshots of the deformed structures. As the schwarzites are very porous and elastic structures, a nature deformation mechanism is 'pore' closing \cite{woellner_MRS_schwarzites_2018,sajadi_2018,felix_2019,felix_2020}, which are present for all structures shown in Fig. \ref{fig:stress}. But only G668 exhibits a pore collapse (indicated the abrupt drops in the stress-strain curves. The whole process can be better understood from the videos1-3 in the Supplementary Materials. This behavior was also reported for other carbon-based schwarzites \cite{felix_2019}.

\section{Summary and Conclusions}
In summary, we have investigated using first-principles methods (SIESTA) the structural stability, mechanical and optical properties of different families of germanium-based schwarzites.

Our results show that all structures are stable and with formation energy values compatibles with other previously syntheszed germanium allotropes (such as germanenes).
The obtained optimized structures are structurally very similar to the ones reported to carbon-based schwarzites. Results from \textit{ab initio} molecular dynamics simulations at $T=300$ K, do not indicate significant thermal structural changes.

Germanium-based schwarzites can be metallic (D688) or small bandgap semiconductors ($0.27$ and $0.13$~eV for G688 and P8bal structures, respectively). 

They can absorb light in a large spectrum, from infrared to ultraviolet and with average reflectivity around about 40-50\% and refractivity index about 2-3 in the visible region. 
These results  suggest that these structures can be good candidates for building optoelectronics devices, solar cell applications, and ultraviolet blockers. 

One important issue is the synthesis feasibility of these structures. The schwarzite synthesis has been elusive, but there are recent important synthesis advances exploiting zeolite as templates \cite{nishihara_2009,nishihara_2018,braun_2018}. It is expected that schwarzite large-size fragments could be a reality in the coming years. We hope the present study can stimulate further studies on these remarkable germanium allotrope structures.

\section{Acknowledgements}
This work was financed in part by the Coordenação de Aperfeiçoamento de Pessoal de Nível Superior - Brasil (CAPES) - Finance Code 001 and CNPq and FAPESP. The authors thank the Center for Computational Engineering and Sciences at Unicamp for financial support through the FAPESP/CEPID Grant \#2013/08293-7.

\bibliographystyle{unsrt}

\end{document}